\documentclass[jcp,aps,final,preprint,groupedaddress,floatfix,aip,reprint]{revtex4}
\input pstricks \input epsf
\usepackage{graphicx}

\begin{document}

\def\beq{\begin{equation}}
\def\eeq{\end{equation}}


\title{Molecular dynamics simulations of the ice temperature dependence of water ice photodesorption}

\author{C. Arasa}
\affiliation{Leiden Observatory, Leiden University, P. O. Box 9513, 2300 RA Leiden, The Netherlands and \\ Gorlaeus Laboratories, Leiden Institute of Chemistry, Leiden University, P. O. Box 9502, 2300 RA Leiden, The Netherlands}

\author{S. Andersson}
\affiliation{Leiden Observatory, Leiden University, P. O. Box 9513, 2300 RA Leiden, The Netherlands and \\
 Gorlaeus Laboratories, Leiden Institute of Chemistry, Leiden University, P. O. Box 9502, 2300 RA Leiden, The Netherlands and \\
 SINTEF Materials and Chemistry, 7465 Trondheim, Norway}

\author{H. M. Cuppen}
\affiliation{Leiden Observatory, Leiden University, P. O. Box 9513, 2300 RA Leiden, The Netherlands}

\author{E. F. van Dishoeck}
\affiliation{Leiden Observatory, Leiden University, P. O.  Box 9513, 2300 RA Leiden, The Netherlands}

\author{G.-J. Kroes}
\affiliation{Gorlaeus Laboratories, Leiden Institute of Chemistry, Leiden University, P. O. Box 9502, 2300 RA Leiden, The Netherlands}

\date{\today}
             

\begin{abstract}
The  ultraviolet (UV) photodissociation of amorphous water ice at different ice temperatures is investigated using Molecular Dynamics (MD) simulations and analytical potentials. Previous MD calculations of UV photodissociation of amorphous and crystalline water ice at 10~K [S. Andersson $et~al.$, J. Chem. Phys. {\bf{124}}, 064715 (2006)] revealed -- for both types of ice -- that H atom, OH, and H$_{2}$O desorption are the most important processes after photoexcitation in the uppermost layers of the ice. Water desorption takes place either by direct desorption of recombined water, or  when, after dissociation, an H atom transfers part of its kinetic energy to one of the surrounding water molecules which is thereby kicked out from the ice.  We present results of MD simulations of UV photodissociation of amorphous ice at 10, 20, 30, and 90~K in order to analyse the effect of ice temperature on UV photodissociation processes. Desorption and trapping probabilities are calculated for photoexcitation of H$_{2}$O in the top four monolayers and the main conclusions are in agreement with the 10~K results: desorption dominates in the top layers, while trapping occurs deeper in the ice. The hydrogen atom photodesorption probability does not depend on ice temperature, but OH and H$_{2}$O photodesorption probabilities tend to increase slightly ($\sim$~30~$\%$) with ice temperature. 
We have compared the total photodesorption probability (OH + H$_{2}$O) with the experimental total photodesorption yield, and in both cases the probabilities rise smoothly with ice temperature.   
The experimental yield is on average  3.8 times larger than our theoretical results, which can be explained by the different time scales studied and the approximations in our model. 

\end{abstract}

\maketitle 

\section{Introduction}

In interstellar space the formation of many simple and complex molecules is mainly driven by reactions on the surfaces of nano- to micrometer sized particles \cite{Herbst2009}. These interstellar grains consist of a core of amorphous silicates or carbonaceous material and, in the cold inner parts of interstellar clouds, they are covered by a mantle of ice, mainly consisting of H$_{2}$O but in many cases also containing large amounts of CO, CO$_{2}$, NH$_{3}$, and CH$_{4}$, among others \cite{Tielens1991, Boogert2008}. Infrared (IR) spectra have revealed that water and carbon monoxide are the most abundant molecules in these icy mantles \cite{Tielens1991, Tanaka1994, Chiar1995, Gibb2000, Gibb2004, Pontoppidan2003, Pontoppidan2006,  Boogert2008, Zasowski2009}. Ultraviolet (UV) irradiation of an ice-coated interstellar grain may cause evaporation of the ice and can drive the formation of complex molecules through the release of reactive radicals by photodissociation even when the flux of UV photons is quite low  \cite{Herbst2009, Fluxphotons, Garrod2006}. Photon fluxes can be of the order of 10$^{3}$ photons cm$^{-2}$s$^{-1}$, which corresponds to roughly  one incident photon per month per grain, while the photodissociation dynamics take place on the picosecond time scale. On this time scale, photodissociation by one incident photon is complete by the time the next photon arrives and the products have completely thermalized. The energies of the available photons,    ranging from $\sim$~6--13~eV \cite{Kobayashi1983, vanDishoeck1988, Boogert2004},  cover the first absorption band of ice.

The study of UV photodissociation and photodesorption of water ice is necessary in order to understand astronomical observations  of gas-phase water and related species in cold clouds \cite{Jack1988, Knacke1991, Cernicharo1990, Gensheimer1996, vanderTak2006, Kaufman2008, Hollenbach2004, Hollenbach2009}. 
UV photodissociation can also lead to the formation of energetic products, in particular energetic H atoms and OH radicals that can move through the ice, encounter other species, and eventually undergo reactions that may lead to the formation of more complex molecules \cite{dHendecourt1982, Garrod2006}. 

Several experiments on UV irradiation of water ice have been carried out using different analysis techniques and different light sources \cite{Westley1995a, Yabushita1, Oberg1}. Ice may also be prepared in different ways, like vapor deposition \cite{Westley1995a, Yabushita1} and under ultrahigh vacuum conditions on a gold substrate \cite{Oberg1}.
Westley $et~al.$ \cite{Westley1995a, Westley1995b} determined the photodesorption yield of H$_{2}$O ice to be (3--8)~$\times$~10$^{-3}$ molecules per UV photon for a 500~nm thick H$_{2}$O ice using a photon source consisting mostly of Lyman-$\alpha$, calibrated UV detectors, a UV-visible spectrometer, and a quadrupole mass spectrometer (QMS). In their experiment the photodesorption rate depends on ice temperature  as well as UV fluence. 
{\"O}berg $et~al.$ \cite{Oberg1} detected  water photodesorption from amorphous ices (at 18--100~K) directly upon  exposure to a UV lamp (7--10.5~eV). The experimental setup used by {\"O}berg $et~al.$ \cite{Oberg1} allowed simultaneous detection of molecules in the gas-phase by QMS and in the ice by reflection absorption infrared spectroscopy (RAIRS). Consistent with these measurements, Yabushita $et~al.$ \cite{Yabushita1} also detected direct desorption of H$_{2}$O ($v$=0) from amorphous solid water and polycrystalline ice after irradiating the ice at $\lambda$=157~nm at 90~K.
{\"O}berg $et~al.$ did not observe a dependence on photon fluence, and attributed the finding by Westley $et~al.$ to H$_{2}$O freeze-out during the early stages of their experiments.

Yabushita $et~al.$  observed H atoms \cite{Yabushita2006} and H$_{2}$ molecules \cite{Yabushita2008} photodesorbing from amorphous water ice while they  irradiated the ice at $\lambda$=157~nm and $\lambda$=193~nm  at 100~K. Using time of flight mass spectroscopy, they observed that photodesorption of H atoms at 100~K depends on the morphology of the ice (amorphous or crystalline). 
This is in contrast with the data of Westley $et~al.$ \cite{Westley1995a} and also with the theoretical results of Andersson $et~al.$ \cite{Andersson2006}. In the latter work it was found that the probabilities of the outcomes and the trends are quite similar for both ice surfaces, although the probability of H photodesorption is higher  for the amorphous  than for the crystalline  ice (e.g., 18$\%$ higher for the first monolayer). 
The experimental results by Yabushita $et~al.$ \cite{Yabushita2006} showed that a large fraction of the H atoms desorbing from amorphous ice had kinetic energies equal to the thermal energy. This suggests that these atoms are released in pores within the ice and do not desorb until 
thermalization occurs through collisions with the  surfaces of the pores. In the theoretical  simulations by Andersson $et~al.$ \cite{Andersson2006, Andersson2008} the model amorphous ice surface is compact, i.e., without large pores, which could explain the difference with experiments.  Another possible source of discrepancy is ice temperature, since all simulations so far refer to 10~K instead of 100~K.

In other experiments, \cite{Ghormley1971} H, OH, and H$_{2}$O$_{2}$ were also detected after flash photolysis of crystalline ice at 263~K. Gerakines $et~al.$ \cite{Gerakines1996} observed OH, HO$_{2}$, and H$_{2}$O$_{2}$ after UV irradiation of amorphous ice at low ice temperature (10~K). Watanabe $et~al.$ \cite{Watanabe2000} irradiated amorphous D$_{2}$O ice at 12~K with 126~nm (9.8~eV) and 172~nm (7.2~eV) light. D$_{2}$ was mainly detected after 126~nm irradiation.  {\"O}berg $et~al.$ \cite{Oberg1} detected OH, H$_{2}$O, H$_{2}$, and O$_{2}$ from amorphous ice at 18--100~K after UV irradiation in the range of 7.0--10.5~eV. They observed that the photodesorption yields tend to increase slightly with ice temperature. Hama $et~al.$ studied vacuum ultraviolet photolysis of amorphous water ice with 157~nm light at 90~K and  detected desorbing species such as OH \cite{HamaAug2009}, O($^{3}$P) \cite{HamaSet12009}, and O($^{1}$D) \cite{HamaSet22009} after irradiation. 
Our study focuses on the dynamics of a single photoexcitation in the uppermost layers of the ice within the picosecond time scale. Therefore any products other than those resulting directly from the water ice photoexcitation  are beyond the scope of this paper.
In all the experiments mentioned above, the crystalline and amorphous ice surfaces are studied by using flat ice samples with dimensions in the cm range, whereas in the interstellar medium (ISM) the ice is frozen onto silicate and/or carbonaceous cores of particles with a typical size of 0.1~$\mu$m.

The photodissociation of water molecules in amorphous and crystalline ice has  been simulated previously using Molecular Dynamics (MD) calculations \cite{Allen1987} at 10~K \cite{Andersson2005, Andersson2006, Andersson2008}. The results indicate that upon absorption of UV photons water molecules and/or H atoms and OH radicals may desorb if the absorption occurs in the outermost layers of the ice. The desorption of water constitutes a relatively unlikely event, whereas desorption of H atoms has a high probability. Deeper in the ice the H and OH released following photodissociation are about equally likely to recombine or to be trapped at separate locations in the ice. In most cases H atoms travel  5--10~\AA~and OH radicals  travel up to 5~\AA~from their original locations, but some H and OH can travel up to tens of~\AA, especially at the ice surface. Mobility could become higher at increased temperatures.
This could have profound  implications for the likelihood of their reactions with other species in the ice. In this paper, we present results for molecular dynamics simulations of the photodissociation of H$_{2}$O molecules in amorphous ice at  astrochemically relevant ice temperatures (10--90~K). Thus, we have investigated the influence of ice temperature on the photodesorption dynamics.

This paper is organized as follows: In Sec.~\ref{sec:methods} we summarize the methods used in this study and describe how the amorphous ice was built. Our results are presented and discussed in Sec.~\ref{sec:results} and we present our conclusions in Sec.~\ref{sec:conclusions}.

\section{Methods}
\label{sec:methods}
In this section, we present the different computational methods employed, including potentials, how the ice is set up, and how the classical dynamics calculations are carried out.

\subsection{Potentials}
\label{ssec:Potentials}
We first discuss the analytical potentials that are used in order to have a good description of all  the interactions that take place  in and on the ice. The total potential can be written as:\\
\begin{equation}\label{eq1}
V_{\rm{tot}} =V_{\rm{ice}}+V_{\rm{H_{2}O^{*}-ice}}+V_{\rm{H_{2}O^{*}}} 
\end{equation}
The first term describes the intermolecular interactions between the H$_{2}$O molecules inside the ice excluding the H$_{2}$O molecule that will be photo-excited. These interactions are based on the TIP4P potential \cite{TIP4P} which consists of O-O Lennard-Jones interactions and of electrostatic interactions between charges on the H atoms and an additional charge site M near the O atom (H:+0.52$e$, M:-1.04$e$) on different molecules. All molecules inside the ice are kept rigid.  The TIP4P potential was parameterized against liquid water at 298~K \cite{TIP4P}, but has nonetheless been shown to provide a qualitatively correct description of the ice phase diagram \cite{Sanz}, and  has been used in several studies to simulate  water ice \cite{Kroes, Karim, MacBride, Martonak, Matsumoto, Buch1, Buch2, Buch3}. In our previous molecular dynamics studies of photodissociation of water in crystalline and amorphous ices \cite{Andersson2005, Andersson2006, Andersson2008}, the TIP4P potential was used to describe water--water interactions at 10~K. For a discussion on the accuracy of the potential, see the Appendix~\ref{Append:A}.

The second term is the intermolecular interaction between the H$_{2}$O molecule that we choose to be photo-excited (this molecule is not rigid, but fully flexible) and the other molecules inside the ice. This interaction is based on the TIP3P potential \cite{TIP4P} with similar O--O Lennard-Jones parameters to TIP4P, but with atom-centered charges on the H and O atoms (H:+0.417$e$, O:-0.834$e$) instead of charges on the H atoms and M sites. 
For the initial interaction (when the molecule is in its first excited state) the TIP3P charges are switched to (H:+0.1$e$, O:-0.2$e$) \cite{Andersson2006}.
The excitation energy is calculated using the TIP3P potential for the ground-state H$_{2}$O (prior to excitation) and the excited-state TIP3P potential (after excitation) (for more details see Ref.~31).
The second term in Eq.~\ref{eq1} also contains the interaction of the photofragments with the ice, i.e., the H--H$_{2}$O  and  OH--H$_{2}$O interactions. The H--H$_{2}$O potential has  been calculated as a reparametrization of the accurate YZCL2 gas-phase H$_{3}$O potential energy surface (PES) \cite{H-H2O}. 
The OH--H$_{2}$O potential has been  constructed \cite{Andersson2006} in a similar way to the Kroes-Clary potential  for the HCl--H$_{2}$O system  \cite{HCl-H2O}. 
In order to smoothly connect the different parts of the V$_{\rm{H_{2}O^{*}}-ice}$ potential, we have used several switching functions to allow the system to be switched from the photo-excited H$_{2}$O molecule to its photofragments, H and OH. The switching functions are functions of the O--H distances (R$_{\rm{O-H}}$) within the molecule and give the interaction parameters as continuous functions in the range 1.1--1.6~\AA. More details about the construction of the PESs and the analytical expressions of the switching functions are given in our previous study \cite{Andersson2006} and in its supporting material. 

In the dissociated state (R$_{\rm{O-H}}$~$>$~1.6~\AA) the partial charges are those of the H atom (0$e$) and OH (function of the O--H bond distance) and the intermolecular interactions are decided by the H--H$_{2}$O and OH--H$_{2}$O potentials. If the molecule recombines, its intermolecular interactions are switched to those of the TIP3P potential.

Finally, the last term is the intramolecular  potential of the photo-excited H$_{2}$O molecule,  which is described by means of the Dobbyn $\&$ Knowles (DK) potential energy surface \cite{DK} based on high-quality ab initio electronic structure calculations. It  contains the intramolecular interactions for the ground state and for the first excited state of  gas-phase H$_{2}$O.  
The DK PES for the first excited state ($\rm{\tilde{A}^{1}B_{1}}$) is a repulsive potential, that leads  to H$_{2}$O dissociation into H + OH.
When the molecule is nearly dissociated and the first excited state has become nearly degenerate with the ground state (when the O--H distance is between 3 and 3.5~\AA), a smooth switch is made to the ground state DK PES. To smoothly connect the excited and ground states, we use a switching function that connects the states linearly in that region \cite{Andersson2006}. Thus, the switching  starts when the O--H distance is between 3 and 3.5~\AA; if the bond remains shorter than 3~\AA~the system will remain in the excited state, but once it becomes larger than 3.5~\AA~it will be switched to the ground state \cite{Andersson2006}.

To avoid interactions with the periodic image cells, the H$_{2}$O--H$_{2}$O, OH--H$_{2}$O, and H--H$_{2}$O interactions are all set to zero at distances greater than 10~\AA~through cutoff functions \cite{HCl-H2O}.

\subsection{Amorphous ice surface}
\label{ssec:Ice}
Initially, crystalline ice is modeled as  normal hexagonal ice (I$_{\rm{h}}$) and its infinite basal plane (0001) face is constructed by applying  periodic boundary conditions in the $x$ and $y$ directions. Although the primitive unit cell is hexagonal, the simulation box is taken rectangular with  parameters: $a$=22.4~\AA, $b$=23.5~\AA, and $c$=29.3~\AA, where the origin of the $z$ coordinate is chosen to be at   the bottom of the ice surface.
The thickness of the ice is given by the $c$ parameter, where the periodic boundary conditions are not applied. Thus, the thickness of the ice is around 30 times smaller than the thickness of an icy grain in the ISM, but thick enough to be representative of ice photodesorption, because photodesorption mainly takes place in the top three monolayers of the ice \cite{Andersson2006, Andersson2008}. 
The crystalline ice slab consists of eight bilayers (BL) or 16 monolayers (ML) and  each ML contains 30 H$_{2}$O molecules (30 in each ML). All of the  molecules are treated as rigid rotors. The molecules in the top 12~MLs are allowed to move, while those in  the bottom 4~MLs are kept fixed in order to simulate bulk ice. Furthermore, the 480 molecules that comprise the unit cell of the ice obey the ice rules \cite{IceRule}, the unit cell having zero dipole moment. The TIP4P pair potential \cite{TIP4P} is used to describe the interactions between the water molecules. 

To prepare  amorphous ice we employ the `fast quenching' method \cite{Essmann1995, AlHalabi2004a, AlHalabi2004b, Andersson2006, Andersson2008}  to set up  ice surfaces at 10, 20, 30 or 90~K.    First,   the crystalline ice is allowed to equilibrate at 10~K for 5~ps,  then   the ice is heated to 300~K in 10~ps (20~ps when  the ice is prepared at 90~K \cite{AlHalabi2004a, AlHalabi2004b}) using the Berendsen thermostat \cite{thermostat}. At 300~K, the molecules in the top bilayers behave like a liquid. In the third step,  the thermostat is switched off and the system is allowed to equilibrate  at 300~K for 100~ps (120~ps when  the ice is prepared at 90~K). After that time, we cool  the ice   to  the desired  ice temperature  (i.e., 10, 20, 30 or 90~K) for  30~ps (15~ps at 90~K) using the thermostat. Then  the system is equilibrated over 120~ps. The simulations use  a leapfrog algorithm \cite{HCl-H2O} to integrate  Newton's equations of motion. 

The structure of the amorphous ice obtained is close to that of compact amorphous ice obtained experimentally, hence it does not show the  microporous structure that is obtained through the vapor deposition  technique \cite{Mayer1986, Kimmel2001a, Kimmel2001b}. Nevertheless, it is a good representation on a local scale \cite{AlHalabi2004a, AlHalabi2004b} of an amorphous ice surface. Furthermore, the calculated average density of the amorphous ice using this method at 90~K \cite{AlHalabi2004a} is about 0.93~g~cm$^{-3}$, which is in good agreement with the density  obtained experimentally for compact low density amorphous (lda) ice at 70~K \cite{Masuda1998} ($\rho$=0.93~g~cm$^{-3}$).

The resulting amorphous surface also contains an equivalent of 16~MLs, but the molecules do not follow a layered structure and a definition of a monolayer in an amorphous ice has to be chosen. One possible definition of a monolayer is based on the thickness of ice corresponding to half of a crystalline bilayer, although the amorphous ice has an irregular bonding structure at the surface.  There are a few molecules that are only two coordinated (for more details see Ref.~55) 
and the distribution of the centers of mass coordinates is not uniform. In our most straightforward  definition,  molecules are attributed to monolayers based on their  $z$ center of mass coordinates, i.e.,  in the first ML there will be the 30 first molecules with largest $z$ center of mass and so on for the following monolayers (binning method 1) \cite{Andersson2006, Andersson2008}. Using this definition, the defined monolayers do not always trace the interface. Especially for corrugated surfaces, the monolayers defined in this way can be very different from what one intuitively expects. 
We have therefore decided to also test a new definition of monolayer, to reassign molecules in  the same top 4~MLs found from the binning method 1 in a different way.
This definition follows basically a similar concept to the previous definition, but on a more local level. The procedure to determine to which monolayer an individual molecule belongs is as follows: first the seven closest neighbors in terms of ($x$,$y$) coordinates to the selected molecule are determined, taking into account periodic boundary conditions. This leads to eight molecules    ($N_{\rm{local}}$)   that are divided in four bins of two molecules each, based on their   $z$ center of mass coordinates.  We assume that these four bins trace the four MLs. The monolayer to which the selected molecule belongs (1--4) is decided from this binning (binning method 2). The outcome of this procedure was checked visually by looking whether the defined monolayers follow the surface corrugation. Larger values of  $N_{\rm{local}}$  (multiples of four) were tested, as well as ($x$,$y$) distance  cut-off values, but these were found to give visually a poorer binning result.
It should be realized that these are just two of many possible binning methods, and that  for amorphous ice attributing the  molecules  to monolayers will always be affected by some level of arbitrariness.

\subsection{Initial conditions}
\label{ssec:Initial}

To study ice photochemistry by means of MD simulations, we choose one water molecule to be photo-excited. This molecule is treated as completely  flexible and its intermolecular interactions  are based on the first  excited electronic state  gas-phase H$_{2}$O ($\rm{\tilde{A}^{1}B_{1}}$)  DK potential energy surface \cite{DK}, which is fully repulsive. Thus, the absorption (of a UV photon) into this state  leads to the dissociation of an isolated  H$_{2}$O molecule into two photofragments: H and OH. In the gas-phase, the probability for dissociation is 100~$\%$, but in the solid state H and OH can also   recombine quickly on the ground state PES.

The excitation energy is computed by taking the energy difference between an ice slab with an excited water molecule and one with a ground state water molecule (both molecules with the same coordinates). A weight is assigned to each excitation  by  calculating  the square of the coordinate dependent transition dipole  moment \cite{Andersson2008}.  This way, even though the treatment of the excited state is simplified (see Sec.~\ref{ssec:Potentials}  and the Appendix~\ref{Append:A}), the calculated spectra of the first UV absorption band in amorphous and crystalline ice are in good agreement with  experiment \cite{Kobayashi1983}. The calculated amorphous  ice spectrum is blue-shifted with respect to that of gas-phase H$_{2}$O, has a significant cross section in the  range of 7.5--9.5~eV,  and has a peak at 8.6~eV (for more details see fig.~3 in Ref.~31 and fig.~1 in Ref.~32). 

A Wigner phase-space distribution function \cite{Wigner1} fitted to the ground-state vibrational wavefunctions  of gas-phase water is used to initialize the trajectories \cite{Wigner2}. The coordinates and momenta of the atoms from the chosen water molecule are sampled using a Monte Carlo procedure. Then, a Franck-Condon excitation is performed and the system is put in the first electronically excited state, on the DK $\rm{\tilde{A}^{1}B_{1}}$ PES  \cite{DK}.

For each excited molecule in a particular ML,  200 different initial configurations  are selected. Thus, to describe photoexcitation in each monolayer we  set up 6000 initial conditions. Since we are only interested in the photodissociation mechanism in the top four monolayers (see Sec.~\ref{ssec:HandOH}),  the total number of integrated trajectories is 24000 at each ice temperature of interest. 

The molecule that is chosen to be dissociated is taken to be flexible, and is exchanged with another  that was fully rigid before. This means  that  its initial geometry  is taken according to the rigid TIP4P model \cite{TIP4P} and is exchanged with the geometry generated from the sampling. In order to simplify this exchange, the molecular plane, the centers of mass and the bisector of the angle HOH are kept unchanged \cite{Andersson2006, Andersson2008}.

\subsection{Dynamics}
\label{ssec:Dynamics}

 Molecular dynamics simulations have been employed to study the dynamics after the photoexcitation of water molecules in amorphous water ice at 10, 20, 30, and 90~K. We assume that classical dynamics are appropriate to describe these systems, since the excited photofragments are very energetic and we do not expect any quantum effects to be important, especially since most of the energy is in the H atom. Since our simulations are terminated once the hydrogen atom becomes thermalized, we do not reach a situation where tunneling could become dominant. For further discussion on possible quantum effects we refer to the Appendix~\ref{Append:A}.

Once  the 200  configurations are set up for one molecule, we run the molecular dynamics code. In the photo-excited water molecule, an elongation of one of the OH bonds always occurs at first  because of the repulsive DK PES\cite{DK}, and when  dissociation occurs it is fast (around 10~fs). 
The maximum time of the dynamics is 20~ps with a time step of 0.02~fs, but the dynamics end when the trajectories are classified according to one of six outcome criteria.  Trapping is defined to occur when an H atom,  OH radical, or H$_{2}$O molecule is accommodated to the ice surface and  its translational energy equals $k_{B}T$ or lower, while the binding energy to the surface $\geq$~0.02~eV for the H atom, $\geq$~0.1~eV for  OH, and $\geq$~0.3~eV for  H$_{2}$O  in order to guarantee that the atoms cannot escape from the surface.  Desorption of the H$_{2}$O molecule or other photofragments is defined as having occurred when its distance  above the surface reaches 11~\AA~and its velocity points towards the vacuum.

There are six different outcomes/channels after  dissociation: 1) H desorbs while OH is trapped inside the ice, 2) OH desorbs while H is trapped inside the ice, 3) H and OH both desorb, 4) H and OH  are both trapped inside the ice, 5) H and OH recombine and form an H$_{2}$O molecule which either desorbs or 6) is trapped inside the ice. An additional outcome is possible in parallel where  H$_{2}$O desorbs through the so called `kick-out' mechanism \cite{Andersson2006, Andersson2008}.  This occurs  when a surrounding water molecule is  `kicked out' by an energetic H atom released after photodissociation,  through an accompanying transfer of momentum. The molecule  is then classified as  `kicked out'  when  its distance to the surface is larger than
11~\AA~ and its velocity is  positive.
We calculate the probabilities of the outcomes/channels per monolayer.

\section{Results and discussion}
\label{sec:results}

\subsection{H atom and OH photodesorption per monolayer}
\label{ssec:HandOH}
UV photodissociation of crystalline and amorphous water ice has already been studied theoretically at 10~K \cite{Andersson2005, Andersson2006,  Andersson2008}. The results show that H$_{2}$O photoexcitation in both crystalline and amorphous ice mainly leads to the desorption of H while OH is trapped in the ice in the top three monolayers. In the deeper monolayers, the H and OH photofragments either recombine or are trapped in the ice at separate positions.  Closer to the surface the photofragments can also recombine and form a water molecule which desorbs. In most cases H atoms travel 5--10~\AA~and OH up to 5~\AA~from their original location, but sometimes H and OH can  travel tens of~\AA~\cite{Andersson2005, Andersson2006,  Andersson2008}. Our results at 10~K agree with these conclusions, although the photodesorption probabilities calculated in this work are slightly larger (0.07~$\%$) than the previous ones \cite{Andersson2008}. 
The  small differences suggest that  the results depend somewhat on the initial geometry of the amorphous ice surface, which in turn depends strongly on the initial conditions in the set up of the  ice surface. Since the amorphous ice geometry is different for each temperature studied, the oscillations in the probabilities that are observed in the figures presented in the following sections, may be partially due to this difference in geometry. This suggests that the sample of molecules in the simulated ice surface is large enough to give consistent results between different geometries at the same ice temperature, but that the finite sample size still introduces some uncertainty in the results. 
We expect this effect to lead to a maximum uncertainty of 5~$\%$ in the probabilities based on the comparison of the 10~K results.

The  probabilities of all  possible outcomes at different ice temperatures (10, 20, 30 and 90~K) have been calculated, but here we only report  results concerning photodissociation followed by desorption of one of the photofragments, both photofragments, or photodesorption of a water molecule 
 (i.e., outcomes 1, 2, 3, 5) for the top four monolayers, which is mainly where  the photodesorption takes place according to previous  MD results at 10~K~\cite{Andersson2006,  Andersson2008}.
In our simulations, the probabilities are calculated per absorbed photon and not per incident photon as in  experiments when  the photodesorption yield is measured.

\begin{figure}[htbp]
\begin{center}
\includegraphics[width=14cm]{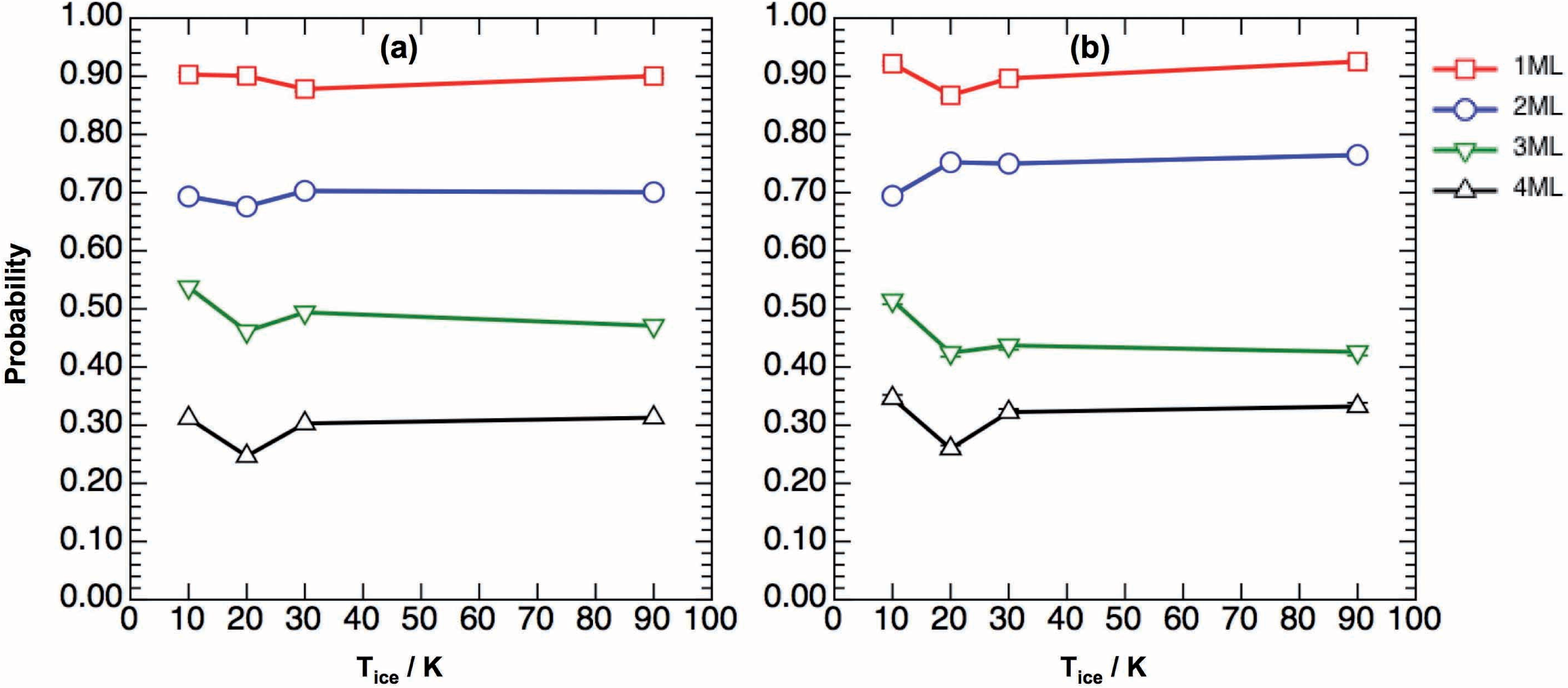}
\end{center}
\caption {{ Total probability of H atom photodesorption (per absorbed UV photon) versus temperature, and for the uppermost four MLs calculated (a) with binning method 1, (b) with binning method 2. }}
\label{Figure1}
\end{figure}

The probabilities of the different outcomes  depend on how deep in the ice  the photo-excited molecule   is initially located. Thus, for excitation in the top monolayers, photodesorption is the most important mechanism, while  deeper in the ice  trapping dominates \cite{Andersson2006, Andersson2008}. The total probability of H atom photodesorption  is calculated by summing over two different processes:  one in which H desorbs while OH is trapped in the ice, and one in which both H and OH desorb from the ice surface (outcome 1 and 3). The H atom desorption probability is much higher than the probabilities of the other events in the top three monolayers at 10~K \cite{Andersson2006, Andersson2008} and  at higher ice temperatures, because in the upper monolayers the structure is more open and the route to the vacuum shorter, which  facilitates the desorption of H atoms after they are  formed. In Figs.~\ref{Figure1}(a) and~\ref{Figure1}(b)  the total probability of H atom desorption versus ice temperature are presented for the top four monolayers (using binning method 1 and 2, respectively), where the  monolayer refers to where the photoexcited H$_{2}$O is located. The dependence on ice temperature and on binning method are negligible, but the dependence on monolayer is rather striking. In the top two monolayers, H desorption is the most important mechanism, while in the third and fourth monolayers, the probability drops because there are other competing pathways, and because the H$_{2}$O molecules in the uppermost layers can prevent the H atom from leaving the surface.

\begin{figure}[htbp]
\begin{center}
\includegraphics[width=14cm]{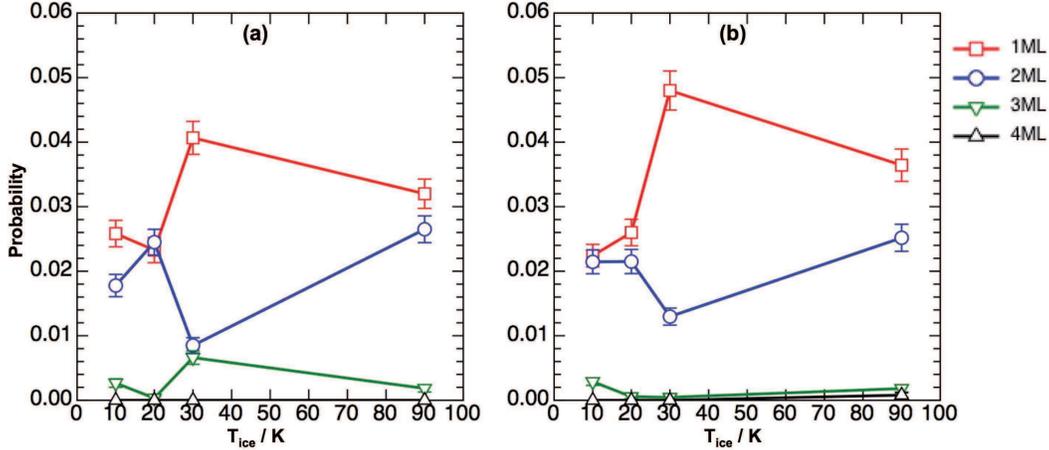}
\end{center}
\caption {{ Total probability of OH photodesorption (per absorbed UV photon) versus ice temperature, and for the uppermost four MLs calculated (a) with binning method 1, (b) with binning method 2. }}
\label{NewFigure1}
\end{figure}

The second most dominant photodesorption mechanism is  OH desorption, which is the sum of two different fractions: a  large fraction contains the probability of channel 3 in which both OH and H  desorb from the surface, and a minor fraction of OH desorbs while H remains trapped in the surface (outcome 2). The total probability depends  on ice temperature (Figs.~\ref{NewFigure1}(a) and~\ref{NewFigure1}(b)), especially  in the top two  monolayers. The OH photodesorption probability decreases  with increasing depth until  the fourth ML is reached,  where the probability is zero for every ice temperature studied, because the OH radicals do not have enough kinetic energy to travel through the top monolayers and escape the surface. 
The OH desorption probability following photoexcitation in monolayers 1--3 shows an oscillatory dependence on ice temperature (Figs.~\ref{NewFigure1}(a) and \ref{NewFigure1}(b) for binnings 1 and 2, respectively). 
We attribute these oscillations to the corrugation of the amorphous ice surface,  which is different for each ice temperature due to the finite simulation cell,  which makes it hard to assign molecules to specific monolayers. 
The finite sample size of about 30 molecules per monolayer can also cause fluctuations in the probabilities.

The H atom desorption probability is more than one order of magnitude larger than the  OH desorption probability. 
This  is because H atoms are lighter and  are formed with higher translational energies  and  lower binding energy to the surface than OH radicals. The abundant H atom desorption is in agreement with the measurements of  Yabushita $et~al.$ \cite{Yabushita2006}, who  measured H atom desorption after UV irradiation of polycrystalline and amorphous ices at 100~K with an excitation energy of 7.9~eV.

\subsection{H$_{2}$O photodesorption per monolayer}
\label{ssec:H2O}

\begin{figure}[htbp]
\begin{center}
\includegraphics[width=17cm]{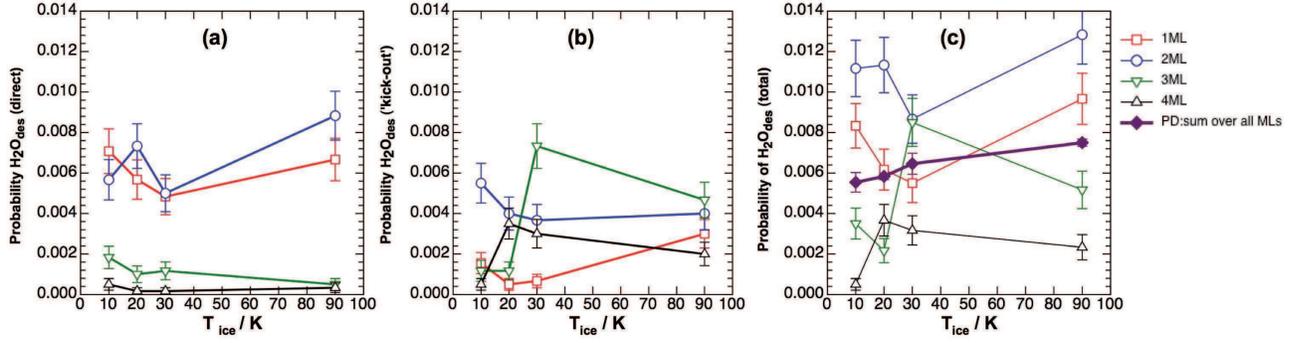}
\end{center}
\caption {{(a) Probability of H$_{2}$O photodesorption due to the direct mechanism, (b) probability of H$_{2}$O photodesorption due to the kick-out mechanism, and (c) total H$_{2}$O photodesorption probability (per absorbed UV photon) versus ice temperature and for the uppermost four MLs calculated with binning method 1.}}
\label{KickoutT}
\end{figure}

\begin{figure}[htbp]
\begin{center}
\includegraphics[width=17cm]{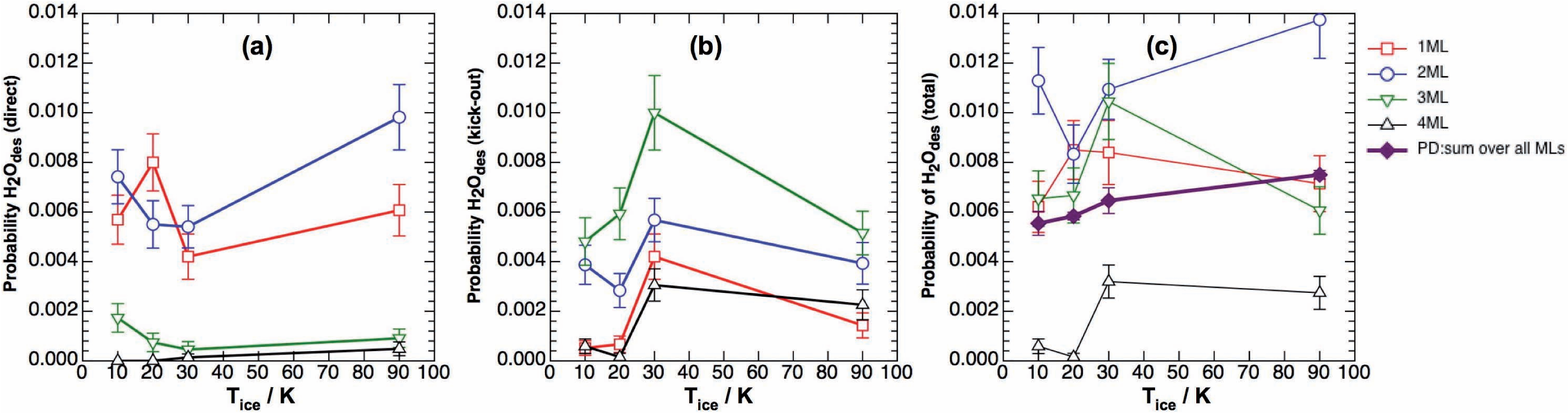}
\end{center}
\caption {{(a) Probability of H$_{2}$O photodesorption due to the direct mechanism, (b) probability of H$_{2}$O photodesorption due to the kick-out mechanism, and (c) total H$_{2}$O photodesorption probability (per absorbed UV photon) versus ice temperature and for the uppermost four MLs calculated with binning method 2.}}
\label{NewKickoutT}
\end{figure}

The third channel of ice photodesorption is H$_{2}$O photodesorption. This is due to two important mechanisms: \\
(1) The recombination of H and OH  gives rise to the formation of H$_{2}$O, which eventually desorbs from the surface (outcome 5). After photoexcitation, H and OH are formed with high translational energies. They may recombine and either desorb as an  H$_{2}$O molecule or become trapped in the ice. If water desorbs, the recombined water molecule that leaves the surface is formed vibrationally excited, according to Andersson $et~al.$ \cite{Andersson2008}. This kind of water photodesorption is direct. The desorption  probability (Figs.~\ref{KickoutT}(a) and~\ref{NewKickoutT}(a)) is small  $<$1~$\%$, and some increase with ice  temperature is only observed in the top two monolayers for $T_{ice}$~$\ge$~30~K. Again, the probability drops when the photoexcitation occurs in the deeper monolayers. 
(2) The second mechanism is due to an energetic H atom formed after H$_{2}$O photodissociation, which transfers momentum to one of the surrounding H$_{2}$O molecules, which eventually desorbs from the ice. This mechanism is hereafter referred to as the kick-out mechanism \cite{Andersson2008}. 
This  photodesorption  mechanism is significant at higher ice  temperatures  and for photoexcitations in  the  second and third monolayers, because  an accelerated atom in these monolayers is more  likely to  kick-out a molecule  above it. This behaviour has been examined by  previous MD calculations at 10~K \cite{Andersson2008}. 
It was   shown  that when a molecule is kicked out from the surface, it is usually due to the  momentum transfer of an energetic H atom (in the direction of the vacuum) to the O of  the departing H$_{2}$O molecule and, occasionally, caused by the vibrationally excited H$_{2}$O molecule formed by recombination. 
A repulsive interaction between the excited molecule and the kicked out molecule can in some cases also contribute to desorption in combination with the momentum transfer from an H atom \cite{Andersson2008}. 
The kicked out molecule is inherently vibrationally cold, because it is treated as internally rigid.

Figs.~\ref{KickoutT}(b) and~\ref{NewKickoutT}(b) show that the water photodesorption probability due to the kick-out mechanism following photoexcitation in monolayer 2 is  higher than the corresponding probability following excitation in monolayer 1.
With binning method 2 (Fig.~\ref{NewKickoutT}(b)) all monolayers follow a more similar trend: 
for all temperatures studied there is a maximum in the probability of  kick-out after photoexcitation  in monolayer 3, with slightly lower probabilities in monolayer 2.
As seen before for OH desorption (Figs.~\ref{NewFigure1}(a) and~\ref{NewFigure1}(b)), the probability for H$_{2}$O desorption through the kick-out mechanism shows an oscillatory dependence on ice temperature.
Once again, we attribute the oscillations to the wavy nature of the amorphous ice surface, which makes it hard to assign molecules to monolayers, and to the finite sample size.

In Fig.~\ref{KickoutT}(c) and Fig.~\ref{NewKickoutT}(c), we show the total  water photodesorption probability averaged over the first four monolayers. The averaging is made in order to have an overview of the total photodesorption probability versus ice temperature and to avoid effects due to the amorphous ice surface corrugation. This yields a smooth curve with the probability slowly increasing with ice temperature ($\sim$30~$\%$ from 10 to 90~K).
Fig.~\ref{KickoutT}(c) and Fig.~\ref{NewKickoutT}(c) also display the total summed H$_{2}$O photodesorption probabilities versus ice temperature for the top four MLs treated individually, which show oscillations for both binning methods.
The slight  increase of the total probabilities with ice temperature occurs because the molecules have higher initial  kinetic energies  at higher ice temperatures,  which  promotes  desorption.

When a water molecule is  kicked out due to the released energy from an H atom, other processes are also taking place.  We have calculated the probabilities of these processes and  conclude that the most important parallel process is that where the H atom that kicks out the H$_{2}$O molecule also desorbs, while the OH is trapped (Table~\ref{Table2}). This behavior is  expected because  the (H$_{\rm{des}}$+OH$_{\rm{trap}}$) mechanism  has the highest probability in the top monolayers of the ice. 
The second important parallel process  is that in which the recombination of the photofragments leads to the formation of water that  is eventually trapped in the ice. 
The recombination is in most cases not causing the kick-out of a molecule, but is rather
caused by the H atom losing kinetic energy in the collision with the H$_{2}$O
molecule. The collision will also change the direction of the momentum of H, quite likely towards the bulk. 
It is interesting to note that on rare occasions the photoexcitation can
lead to the removal of two water molecules from the ice, since there are small
probabilities of simultaneous desorption of the kicked out molecule and either H and
OH or the recombined H$_{2}$O molecule.

\begin{table}[htbp!]
\caption{Probabilities averaged over the top four  monolayers of the outcomes that take place at the same time as the kick-out mechanism, for each ice temperature $^\mathrm{a}$.}
\begin{center}
\begin{tabular} { c || c | c | c | c | c| c} 
\hline
\hline
$T_{\rm{ice}}$  / K & H$_{\rm{des}}$ + OH$_{\rm{trap}}$ & H$_{\rm{des}}$ + OH$_{\rm{des}}$  & H$_{2}$O$_{\rm{des}}$ & H$_{\rm{trap}}$ + OH$_{\rm{trap}}$ & H$_{2}$O$_{\rm{trap}}$ & Others \\ \hline
10 & 0.600 $\pm$ 0.066 & 0.018 $\pm$ 0.018 & 0 & 0.073 $\pm$ 0.035 & 0.255 $\pm$ 0.059 & 0.054 $\pm$ 0.014\\
20    & 0.509 $\pm$ 0.069 & 0  & 0 & 0.151 $\pm$ 0.049 & 0.340 $\pm$ 0.065 & 0 \\  
30    & 0.548 $\pm$ 0.058 & 0.014 $\pm$ 0.014  & 0.014 $\pm$ 0.014 & 0.082 $\pm$ 0.032 & 0.342 $\pm$ 0.056 & 0 \\
90   & 0.576 $\pm$ 0.064 & 0  & 0.051 $\pm$  0.029 & 0.051 $\pm$ 0.029 & 0.322 $\pm$ 0.061 & 0 \\ \hline
\end{tabular}

\begin{list} {} {}
\item $^\mathrm{a}$ Overall probabilities are obtained by multiplying with the probabilities for the kick-out mechanism, see Figs.~\ref{KickoutT}(b) and \ref{NewKickoutT}(b).
\end{list}
\end{center} 
\label{Table2}
\end{table}

\subsection{Energies of the kicked out H$_{2}$O molecules}
 
Experimental measurements of H$_{2}$O ($v$=0) photodesorption from amorphous water ice at 90~K \cite{Yabushita1} showed that the measured translational and rotational energies of H$_{2}$O after photodesorption were in good agreement with those calculated by means of MD simulations for the kick-out mechanism for an amorphous ice surface at 10~K \cite{Yabushita1}. Thus, the kick-out mechanism is likely to be  one of the main photodesorption mechanisms of H$_{2}$O desorption in its vibrational ground state ($v$=0) while the recombination mechanism more likely results in  desorption of H$_{2}$O in vibrationally excited states. 

\begin{figure}[htbp]
\begin{center}
\includegraphics[width=8cm]{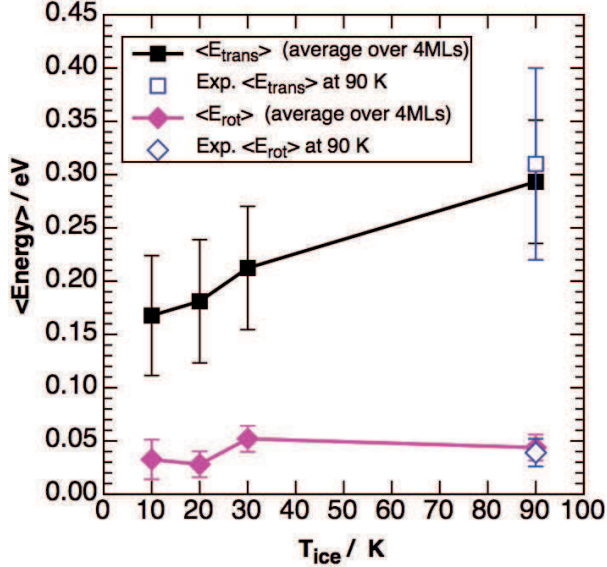}
\end{center}
\caption {{Calculated average translational and  rotational energies  of the kicked out water molecules versus ice temperature, and experimental average translational and rotational energies of  water molecules desorbed in their ground vibrational state at $T_{\rm{ice}}$=90~K \cite{Yabushita1}.}}
\label{Figure3}
\end{figure}

We have calculated the average  translational and rotational energies of the kicked out molecules at the end of each trajectory, and  these energies have been averaged over the top four MLs.  In Fig.~\ref{Figure3} we have plotted the average energies against ice temperature and the corresponding experimental energies at 90~K \cite{Yabushita1}. The translational energy tends to rise with ice temperature. This is due to the energy of the ice (which also rises with ice temperature) being released to the kicked out molecule. The final rotational energy is  low (although higher than thermal) and it shows quite weak  ice temperature dependence in the range between 10 and 90~K. 
The average translational and rotational energy values at 90~K are 0.29 and 0.044~eV, respectively.  That  compares well with the average experimental energies measured by Yabushita $et~al.$ \cite{Yabushita1} at 90~K   (0.31 and 0.039~eV, respectively).
These water molecules leave the surface vibrationally cold ($v$=0), while the molecules that desorb after recombination are formed vibrationally excited.

\subsection{OH and H$_{2}$O photodesorption yields}
\label{ssec:OH+H2O}
We have also calculated the normalised OH desorption product yields (normalised to a OH desorption product yield of 1.0 at 20~K) and the normalised H$_{2}$O desorption product yields (normalised to obtain the computed OH$_{\rm{des}}$/H$_{2}$O$_{\rm{des}}$ ratio at 20~K with an OH$_{\rm{des}}$ product yield of one)   (see Table~\ref{Table1}) to compare better to experiments.

As seen in Table~\ref{Table1}, the normalised OH desorption product yield  is larger than the normalised H$_{2}$O desorption product yield at all ice temperatures in our simulations, whereas experimentally it is the other way around for $T_{\rm{ice}}$~$\ge$~30~K. 
This discrepancy may be due to the larger time scales considered in the experiments.
A stronger temperature dependence would be expected if longer time scales are considered. 
At the short time scales covered in our simulations, thermally activated processes such as diffusion and thermal desorption are not probed since they require much longer time scales. 
Mobile H and OH, formed in the same or in subsequent photoexcitation events, can reach each other, recombine and desorb as a consequence of the excess energy. These secondary processes are beyond the scope of our simulations. Experimental desorption rates are however the result of both the short time scale processes presented here and the long time scale thermal effects. 
The fact that we do not observe a large thermal effect for the short time scale processes indicates that the increase in desorption rate as observed experimentally is  due to the dominance of thermal effects as hypothesized by {\"O}berg $et~al.$ \cite{Oberg1}

\begin{table}[htbp]
\caption{Experimental and theoretical (summed over the top four MLs) desorption product yields  of OH and H$_{2}$O versus ice temperature normalized to a OH desorption product yield of 1.0 at 20 K,  and product yield fractions.}
\begin{center}
\begin{tabular} { c | c c  c  || c | c c c } 
\hline
\hline
$T_{\rm{ice}}$ / K &  &  Experiment \cite{Oberg1}&  & $T_{\rm{ice}}$ / K & & Theory (this work)&  \\ 
 & OH$_{\rm{des}}$  &H$_{2}$O$_{\rm{des}}$  & OH$_{\rm{des}}$/H$_{2}$O$_{\rm{des}}$  &  & OH$_{\rm{des}}$&H$_{2}$O$_{\rm{des}}$ &OH$_{\rm{des}}$/H$_{2}$O$_{\rm{des}}$\\ \hline
      &          &      &   & 10 & 0.95 &0.45& 2.11 \\
20    & 1.00&0.70& 1.43 &     20 & 1.00 & 0.49& 2.04 \\  
30    & 1.20&1.40& 0.86 &     30 & 1.15 & 0.54& 2.13 \\
100   & 1.20&2.00& 0.60 &     90 & 1.26 & 0.63& 2.00 \\ \hline
\end{tabular}
\end{center} 

\label{Table1}
\end{table}

\subsection{Total (OH + H$_{2}$O) photodesorption yield}

\begin{figure}[htbp]
\begin{center}
\includegraphics[width=8cm]{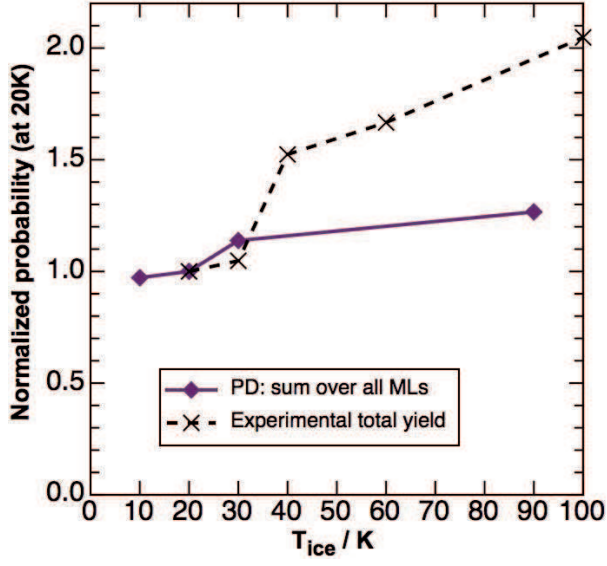}
\end{center}
\caption {{Calculated (solid line) total (OH$_{\rm{des}}$ + H$_{2}$O$_{\rm{des}}$) desorption probability summed over all MLs and the experimental total photodesorption yield (dashed line) \cite{Oberg1}.}}
\label{Figure2}
\end{figure}

Fig.~\ref{Figure2} shows the total photodesorption (OH$+$H$_{2}$O) yield versus ice temperature. These two photodesorption yields are added to compare them directly to experiments where the water photodesorption yields are measured as the combined contributions of H$_{2}$O and OH photodesorption yields.
To calculate the total photodesorption yield, we have summed  the total OH photodesorption and the total H$_{2}$O photodesorption probabilities over the top four monolayers, but we have not added  the H photodesorption probability because it was not measured in the experiments \cite{Oberg1}. Thus, in Fig.~\ref{Figure2} we have plotted the total (OH$+$H$_{2}$O) desorption probability calculated from our simulations (solid line) and that from the experimental measurements (dashed line). Both are normalized  to 1.0 at 20~K because here we want to focus on the dependence of the yield on ice temperature.
Below,  we will compare the experimental and the theoretical data quantitatively, using the same units (photon$^{-1}$).  
Fig.~\ref{Figure2} shows that in both cases the ice temperature dependence is clear, but not dramatic. There is a steady increase of the desorption with increasing ice temperature by at most a factor of two.
The experimental normalized photodesorption yield is larger than our results at high $T_{\rm{ice}}$. 
Again, we attribute the higher experimental yield at high ice temperatures to the contribution of long time scale processes, such as thermal diffusion of 
fragments followed by recombination and subsequent desorption, and thermal desorption (see Sec.~\ref{ssec:OH+H2O}).

Up to this point, all theoretical probabilities have been calculated per absorbed UV photon in a specific monolayer of the ice surface. However, not all of the UV photons that arrive at the ice are absorbed in these monolayers. In our previous paper \cite{Andersson2008}, we  estimated the absorption probability per monolayer. This probability  ($P_{\rm{abs}}^{\rm{ML}}$) is calculated by dividing the cross section for photodissociation by the effective area of a water molecule in a monolayer. Mason $et~al.$ \cite{Mason2006} measured the absorption cross section of water ice at 25~K to be about 6$\times$10$^{-18}$~cm$^{2}$ near an excitation energy of 8.61~eV, from which  the effective area of a molecule in an ML is calculated.  
The absorption probability per monolayer is estimated   around 7$\times$10$^{-3}$ (for more details see Appendix~A in our previous study \cite{Andersson2008}).
From the calculated  probabilities per absorbed UV photon per monolayer, we can then  estimate the photodesorption probabilities per incident UV photon.  
The total photodesorption yield can be calculated by multiplying the total photodesorption probability per absorbed photon in a given monolayer $i$, $P_{\rm{des}}^{i}$, with the probability that the photon makes it to monolayer $i$ and the probability that the photon is absorbed in a given monolayer   ($P_{\rm{abs}}^{\rm{ML}}$), and summing the resulting yields per monolayer over the monolayers. This may be summarised through the equation:

\begin{equation}\label{eq2}
P_{\rm{photon^{-1}}}=\sum_{i=1}^{n}P_{\rm{des}}^{i}\cdot(1-P_{\rm{abs}}^{\rm{ML}})^{i-1}\cdot P_{\rm{abs}}^{\rm{ML}}
\end{equation}\\

Table~\ref{Table3} contains the theoretical and experimental photodesorption yields (in photon$^{-1}$), and also the fraction experimental/theoretical yield at all ice temperatures.
Given the experimental uncertainty (60$\%$) and our approximations (such as the use of gas-phase PESs for the H$_{2}$O intramolecular interactions,   the freezing of  the intramolecular degrees of freedom of the surrounding H$_{2}$O molecules, and the short time scale of our simulations),  the data are considered to be in reasonable agreement.  Our results also agree with the photodesorption probabilities per incident photon range (1$\times$10$^{-4}$--3.5$\times$10$^{-3}$) \cite{Bergin1995, Willacy200, Snell2005, Dominik2005, Bergin2005} used to model astrophysical environments.

\begin{table}[htbp]
\caption{Experimental (calculated from the empirical fitting of the total photodesorption yield, eq.~4 in Ref.~27) and theoretical (OH$_{\rm{des}}$ + H$_{2}$O$_{\rm{des}}$) photodesorption yields (photon$^{-1}$) and the fraction experimental/theoretical yield at all ice temperatures.}
\begin{center}
\begin{tabular} { c   || c | c | c } 
\hline
\hline
$T_{\rm{ice}}$ / K & Exp. yield \cite{Oberg1} / photon$^{-1}$  & Theo. yield (this work) / photon$^{-1}$ & Exp./Theo. yield\\ \hline
10   & (1.62 $\pm$ 0.48)$\times$10$^{-3}$ & (5.38 $\pm$ 0.41)$\times$10$^{-4}$ & 3.0   \\
20    & (1.94 $\pm$ 0.56)$\times$10$^{-3}$ & (5.66 $\pm$ 0.42)$\times$10$^{-4}$ & 3.4   \\  
30    & (2.26 $\pm$ 0.64)$\times$10$^{-3}$ & (7.13 $\pm$ 0.87)$\times$10$^{-4}$  & 3.2  \\
90   & (4.18 $\pm$ 1.1)$\times$10$^{-3}$ & (7.11 $\pm$ 0.47)$\times$10$^{-4}$ &  5.9  \\ \hline
\end{tabular}
\end{center} 

\label{Table3}
\end{table}

The experimental photodesorption yield is larger than our values at all ice temperatures. The ratio experimental/theoretical  yield increases from 3.0 to 5.9 going from 10~K to 90~K. This trend can again (Sec.~\ref{ssec:OH+H2O}) be attributed to long time scale processes becoming increasingly important with increasing ice temperature in the experiments. Another potential source of discrepancy is due to the UV wavelength covered by the lamp  used in the experiments \cite{Oberg1}. This UV   lamp includes Lyman-$\alpha$ photons which can excite the $\rm{\tilde{B}}$ excited state of H$_{2}$O whereas our calculations consider only the $\rm{\tilde{A}}$ state (see also discussion in Ref.~32).

Some OH photofragments and recombined molecules that are trapped in the ice in our simulations    could desorb at long time scales.

The mechanism in competition with  photodesorption is trapping.  Trapping mainly occurs in the deeper monolayers when OH and/or H are trapped in the ice, or when recombination of the photofragments produces a water molecule that remains inside the ice.
The H atom, the OH radical, and the recombined H$_{2}$O molecules move inside the ice until they become trapped. 
The mobility of the H and OH photoproducts is not affected by ice  temperature because the translational energies following photodissociation are orders of magnitude larger than the thermal energies. 
At all temperatures studied, the H atoms produced after water photodissociation move on average  8~\AA~until they are trapped, and the distance traveled by the OH radicals is on average around to 1--2~\AA, while the recombined water moves on average a distance of 2~\AA \cite{Andersson2006}. 
The maximum  distance travelled before trapping is of the order of tens of~\AA. These large  distances make  reactions with other species possible, which could in turn lead to the formation of more complex molecules if species other than H$_{2}$O are present in the ice.

\section{Conclusions}
\label{sec:conclusions}

The primary goal of this work was to investigate the effect of ice temperature on the UV photodesorption of water ice. 
Hydrogen atom photodesorption is the most important desorption channel  in the uppermost layers of the ice, with a probability of about 2 or 3 orders of magnitude larger than those of desorption of OH and H$_{2}$O, and  it displays  no dependence on ice temperature. The second most important desorption mechanism is OH photodesorption. We found that OH only desorbs from the top three monolayers.
H$_{2}$O photodesorption is also observed upon UV absorption in the top monolayers,  either by direct desorption after the recombination of the photofragments or by indirect desorption due to an energetic H atom released by photodissociation that kicks out a surrounding water molecule. The kick-out mechanism mainly takes place after photoexcitation in the second and third monolayers of the ice because a molecule  photo-excited  in these layers will produce an H atom with high kinetic energy that can  easily transfer its momentum to a  molecule in the top layer which is then free to leave the surface. The total water desorption probability   increases by only $\sim$30~$\%$ with increased ice temperature from 10 to 90~K.

The OH photodesorption probabilities and the kick-out H$_{2}$O photodesorption probabilities versus ice temperature  for the top four monolayers show some oscillations due to the corrugation of the amorphous ice surface and to the finite sample size of the simulated ice of about 30 molecules per monolayer, irrespective of the binning method used. This confirms that one has to be careful when assigning molecules to monolayers for amorphous ices where the surface is very  corrugated. 

We have also estimated the total photodesorption probability (OH + H$_{2}$O) per incident photon from our probabilities (calculated per absorbed UV photon) and we have compared our results with the available experimental (OH + H$_{2}$O)  photodesorption yields. Both data rise with ice temperature. The experimental yields are a factor  3.0--5.9 larger than our probabilities. Given the experimental uncertainties  and our approximations (such as the use of gas-phase PESs for the H$_{2}$O intramolecular interactions,  freezing the intramolecular degrees of freedom of the surrounding H$_{2}$O molecules,  using a short time scale, and exciting only the $\rm{\tilde{A}}$ state), we can conclude that our simulations agree reasonably  well with the experimental photodesorption yields, which are also typically used in modelling astrophysical environments.

Both our calculations and the experiments allow us to conclude that  the  dependence of desorption probability on ice temperature is  weak, at most a factor of two, probably because the small thermal energies do not influence much the direct processes occurring after photodissociation which involve released energies of several eV \cite{Andersson2008}. 
Although the agreement with the experiments is reasonable, comparison with the experiments is not straightforward because of different time scales. The time scale in the simulations is picoseconds, while in the laboratory it can be hours, making  processes such as thermal diffusion and thermal desorption increasingly important, especially at high ice temperatures. The better agreement between theory and experiment at low ice temperatures (below 30 K) indicates that at these ice temperatures the short time scales processes have a higher relative importance.

We have shown in our previous papers that upon UV photodesorption in the deeper monolayers of water ice,  trapping is the dominant mechanism, as  H and OH either are trapped as separated fragments at different positions or  as recombined H$_{2}$O  in mixed interstellar ices. The high mobility of H and OH can lead to reactions with coadsorbed molecules and eventually to the formation of more complex molecules, such as CO$_{2}$  if the OH radical  reacts with  coadsorbed CO. Once CO$_{2}$ is formed, it could either desorb if it has enough energy to leave from the ice surface or could stay trapped in the ice.  Thus, the fraction of H atoms and OH radicals that desorb from the ice surface might be different depending on the composition of the ice. But since H$_{2}$O is the main component of interstellar ices, we still expect 
that H photodesorption would be the most important photodesorption outcome in the top three monolayers of the ice, and that water photofragments dominate the photochemistry in the ice.
Therefore, the study of such reactions is a logical extension of the present work and may increase our understanding of  the composition of interstellar ices.

\appendix
\section{Accuracy of the model}
\label{Append:A}
In Sec.~III.I of our previous study, Ref.~31, we discussed a number of approximations that might limit the accuracy of our model. We listed the most important approximations as (i) the use of gas-phase PESs for the H$_{2}$O intramolecular interactions, (ii) the use of (nonpolarizable) pair potentials for the intermolecular
interactions, (iii) freezing the intramolecular degrees of freedom of the surrounding H$_{2}$O molecules, (iv) the simplified treatment of recombination, and (v) the use of classical dynamics for nuclear motion. In view of recent work that has been published
since or that we were not aware of at the time, we would like to complement the previous discussion.

The use of the TIP4P potential for studying ice at low temperatures cannot be solely justified based on its construction. However, despite TIP4P being parameterized against the properties of liquid water at 298~K \cite{TIP4P}, it has been widely
used to study various properties of ice \cite{Sanz, Kroes, Karim, MacBride, Martonak, Matsumoto, Buch1, Buch2, Buch3}. 
Its success and usefulness in this respect can be understood by the fact that TIP4P actually gives a qualitatively correct
description of the ice phase diagram \cite{Sanz}, yields stable hexagonal ice at low temperatures \cite{Kroes, Karim}, and gives a reasonable description of the low-energy part of the
phonon spectrum of hexagonal ice \cite{ApA}. The calculated melting temperature of TIP4P hexagonal ice is somewhat low, 232~K at 1~bar, but since we are studying much lower temperatures this need not be any serious concern.

In Sec.~II.B of Ref.~31 it was discussed that the present use of the modified TIP3P partial charges for the interaction of the excited H$_{2}$O molecule with its surrounding molecules is a simplification of the  `real' interaction, which should likely include exchange-repulsion interactions, polarization, and modification of the
intramolecular part of the excited state potential. 
The partial charges were only modified to make the peaks of the simulated and experimental UV spectra of hexagonal ice coincide, at 8.6~eV. Nevertheless, the shape of the spectrum shows
quite good agreement with experiment, with almost exactly the same relative intensities as the experimental spectrum around the peak, between 8.3 and 9.0~eV. The low-energy threshold is also found at the same energy as in experiment, 7.5~eV \cite{Andersson2006, Andersson2008}. 
This is also the case for the UV spectra of liquid water and amorphous ice \cite{Andersson2006, Andersson2008}. Therefore, we have drawn the conclusion that the amount of excess energy that is released in our simulations is correct. Regarding the partitioning of that excess energy, it seems that the desorption energy of H atoms is somewhat high, by about 1~eV  \cite{Andersson2006, Andersson2008}. However, it is not clear
whether this is mainly due to the H atoms being formed too hot or that possible energy dissipation mechanisms are not well treated, e.g., transfer of energy to intramolecular vibrational modes of H$_{2}$O. The good agreement between the simulated
and experimental translational and rotational energies of kicked out H$_{2}$O (this work and Ref.~26) is in this respect remarkable given the simplified treatment of the excited state potential.

Using classical dynamics to study molecular motion will always be an
approximation to the true quantum mechanical treatment. The quantum effects will often become more important at lower temperatures. For a condensed phase system the main quantum effects are delocalization, conservation of zero-point energy, and tunneling. In our model the initial vibrational state of the photo-excited molecule
includes a proper treatment of zero-point motion through the Wigner distribution. As discussed in Ref.~31 the motion of the highly translationally excited H atom should be
well described by classical dynamics, but if the H atom is thermalized it could diffuse
through quantum tunneling between cages in the ice. A recent path-integral  treatment of H atom diffusion through hexagonal ice at 8~K \cite{ApB} confirms that tunneling indeed is important in this process, but also indicates that thermal diffusion is quite slow at this temperature, $k_{diff}$ = 2.7$\times$10$^{-4}$s$^{-1}$. Experimental evidence and transition-state theory calculations suggest that H diffusion becomes more rapid, by several orders of magnitude, above 50~K (see Ref.~73 and references therein). This could therefore strongly affect the comparison between
our simulations and experiments at 90~K, assuming similar behavior in
amorphous ice.

The quantum effects in pure ice have recently been studied by several groups \cite{ApC,ApD,ApE,ApF}. It was found that clear quantum behavior shows up, mainly due to quantum delocalization, when comparing to classical simulations using the same
potentials. These effects include lowering the melting temperature \cite{ApC,ApD, ApE} enhanced surface premelting  at temperatures just below the melting temperature \cite{ApC, ApD}, reducing the density of ices at all temperatures and changing the relative stability of some phases of ice \cite{ApF}.
Since we are studying quite low temperatures, 10--90~K, the first two of these effects  should not be important, but the others could be.
However, an important observation is that if one uses a water potential parameterized using classical dynamics, e.g., TIP4P, quantum effects will be implicitly included in the potential. Using such a potential when treating the nuclear motion quantum mechanically will most likely give worse agreement
with experiments than a corresponding classical study \cite{ApF}. Therefore, if quantum effects are included in the dynamics, potentials optimized for use with quantum dynamics or polarizable force fields fitted to ab initio data should be used \cite{ApC, ApD, ApE, ApF}. In a classical MD study more reliable results will conversely be obtained with a potential including quantum effects, like TIP4P or similar, regardless of temperature.

Explicit inclusion of additional quantum effects in our study is non-trivial. A possible approach would be to still run classical dynamics, but to extend the initial Wigner distribution to cover more molecules in the ice, i.e., treating more molecules
as flexible and thereby including zero-point motion for these molecules. This has been done in studies of liquid water by the classical Wigner, or LSC-IVR, method \cite{ApG,ApH}.
However, it was found that this approach leads to unphysical leakage of the zero point energy from intramolecular to intermolecular modes, thereby leading to considerable heating of the center-of-mass motion of H$_{2}$O, from 321~K to 700~K, over the course of 1~ps \cite{ApH}. Similar problems would be expected to occur for simulations of ice.

\begin{acknowledgments}

 The authors would like to thank  Dr. T. P. M. Goumans for valuable discussions. 
 This project was funded with computer time by NCF/NWO, and with a TOP grant No. 700.56.321 by CW/NWO.
\end{acknowledgments}


\end{document}